\documentclass[12pt]{article}
\usepackage[dvips]{graphicx}
\usepackage[dvips]{graphics}

 \def\erm{\mbox{e}}
 \def\drm{\mbox{d}}
\newcommand{\newc}{\newcommand}
\newc{\lra}{\leftrightarrow}
\newc{\beq}{\begin{equation}}
\newc{\eeq}{\end{equation}}
\newc{\barr}{\begin{eqnarray}}
\newc{\earr}{\end{eqnarray}}

\title{The Effect of the Short-Range Correlations on the Generalized Momentum
Distribution in Finite Nuclei}

\author{Ch.~C.~Moustakidis$^1$\footnote{\texttt{e-mail:\, moustaki\,@\,auth.gr}}\,,
P.~Papakonstantinou$^2$\footnote{\texttt{e-mail:\,
panagiota.papakonstantinou\,@\,physik.tu-darmstadt.de}}\,, and
E.~Mavrommatis$^3$\footnote{\texttt{e-mail:\,
emavrom\,@\,phys.uoa.gr}}
\\
{\it  $^{1}$Department of Theoretical Physics,} \\
{\it Aristotle University of Thessaloniki,} \\
{\it 54124 Thessaloniki Greece}\\
{\it $^{2}$Institute of Nuclear Physics, } \\
{\it T.U.Darmstadt. Schlossgartenstr.9,}\\
{\it D-64289 Darmstadt, Germany} \\
{\it $^{3}$ University of Athens, Physics Department,}\\
{\it Nuclear and Particle Physics Division,}\\
{\it Panepistimiopolis, Ilissia, GR-15771 Athens, Greece } }

\date{}

\begin{document}

\maketitle

\begin{abstract}
The effect of dynamical short-range correlations on the
generalized momentum distribution $n(\vec{p},\vec{Q})$
 in the case of $Z=N$,
$\ell$-closed shell nuclei is investigated by introducing
Jastrow-type correlations in the harmonic-oscillator model. First,
a low order approximation is considered and applied to the nucleus
$^4$He. Compact analytical expressions are derived and numerical
results are presented and the effect of center-of-mass corrections
is estimated. Next, an approximation is proposed for $n(\vec{p},
\vec{Q})$ of heavier nuclei, that uses the above correlated
$n(\vec{p},\vec{Q})$ of $^4$He. Results are presented for the
nucleus $^{16}$O. It is found that the effect of short-range
correlations is significant for rather large values of the momenta
$p$ and/or $Q$ and should be included, along with center of mass
corrections for light nuclei, in a reliable evaluation of
$n(\vec{p},\vec{ Q})$ in the whole domain of $p$ and $Q$.

\end{abstract}

\

\section{Introduction}

In a system of $A$ $(A \geq 2)$ identical particles described by a
unit-normalized state $|\Psi \rangle$, the generalized momentum
distribution $n(\vec{p},\vec{Q})$ (GMD) is defined by
\cite{Ristig-90,Mavro-95,Papa-00}
\begin{equation}
n(\vec{p},\vec{Q})= \mathcal{N} \langle \Psi|\sum_{\vec{k},
s,s',t,t'}a_{\vec{k}+\vec{Q},s',t'}^{\dagger}
a_{\vec{p}-\vec{Q},s,t}^{\dagger} a_{\vec{p},s,t}
a_{\vec{k},s',t'}|\Psi \rangle \ .\label{GMD-1}
\end{equation}
By introducing the density fluctuation operator
$\hat{\rho}_{\vec{Q}}=\sum_{\vec{k},s,t}a_{\vec{k}+\vec{Q},s,t}^{\dagger}a_{\vec{k},s,t}$\\
 and the single-particle momentum distribution $n(\vec{p})$, definition (\ref{GMD-1}) is recast as
\begin{equation}
n(\vec{p},\vec{Q})=\mathcal{N} [\langle \Psi | \hat{\rho}_{\vec{Q}}
\sum_{s,t} a_{\vec{p}-\vec{Q},s,t}^{\dagger} a_{\vec{p},s,t} |
\Psi \rangle-n(\vec{p}) ] \ . \label{GMD-2}
\end{equation}
The role of GMD in final state interactions becomes evident from
this expression, since the first term on the right is a transition
matrix element for scattering a particle out of the orbital
$(\vec{p},s,t)$ into the orbital $(\vec{p}-\vec{Q},s,t)$, this
process being introduced by a spin-isospin independent density
fluctuation of wave vector $\vec{Q}$.

The quantity $n(\vec{p},\vec{Q})$ is connected to the two-body
density matrix (2DM) in momentum space
$n_2(\vec{p}_1,\vec{p}_2;\vec{p}_{1'},\vec{p}_{2'})$ through the
relation
\begin{equation}
n(\vec{p},\vec{Q})=\int
n_2(\vec{p},\vec{k};\vec{p}-\vec{Q},\vec{k}+\vec{Q}) \drm^3 k \ .
\label{GMD-3}
\end{equation}
Introducing the two-body density matrix in the coordinate space
and its half diagonal version
$\rho_{2h}(\vec{r}_1,\vec{r}_2;\vec{r}_{1'})$ and performing a
Fourier transform in the coordinates $\vec{r_1}-\vec{r}_{1'}$ and
$\vec{r}_{1'}-\vec{r_2}$ we obtain for GMD
\begin{equation}
n(\vec{p},\vec{Q})=\frac{1}{(2 \pi)^3} \int
\rho_{2h}(\vec{r}_1,\vec{r}_2;\vec{r}_{1'}) \erm^{-i \vec{p} \cdot
(\vec{r_1}-\vec{r}_{1'})} \erm^{-i \vec{Q} \cdot
(\vec{r}_{1'}-\vec{r_2})} \drm^3r_1 \drm^3r_{1'} \drm^3r_2  \ .
\label{GMD-4}
\end{equation}
We shall take $\hbar=c=1$, so that the momentum has the dimension
of inverse length and $n(\vec{p})$ and the GMD have the dimension
of (length)$^3$. The normalizations adopted here are such that
$\int n_2(\vec{p}_1,\vec{p}_2;\vec{p_1},\vec{p_2}) \drm^3 p_1
\drm^3 p_2 = A(A-1) $ and $\int n(\vec{p}) \drm^3 p = A$.

Regarding the above definitions, in Refs.[1-3] no distinction was
made between ``laboratory" momenta and intrinsic momenta, i.e.,
momenta with respect to the Center of Mass (CM) frame. For
infinite systems, such as considered in Refs.[1,2], CM
correlations are not relevant. For large self-bound systems like
heavy nuclei, they are also not very important. For small systems,
however, like the nucleus $^4$He, one has to be careful with the
definitions and relations given above. In particular, the momenta
in Eqs. (\ref{GMD-1})-(\ref{GMD-4}) should be interpreted as
momenta with respect to the center of mass of the system. We shall
return to this point later.

The generalized momentum distribution $n(\vec{p},\vec{Q})$ has
some important formal properties that result from the
corresponding properties of the 2DM in momentum or coordinate
space. In particular, the sequential relation between the
half-diagonal 2DM and the one-body density matrix (1DM) yields the
relation between the GMD and $n(\vec{p})$, namely
\begin{equation}
n(\vec{p},\vec{Q}=0)=(A-1)n(\vec{p}) \ .  \label{GMD-5}
\end{equation}
In addition time-reversal invariance implies that
$n(\vec{p},\vec{Q})$ is symmetric with respect to the variables
$\vec{p}$ and $\vec{w}$ ($\vec{w}=\vec{p}-\vec{Q}$) namely
\begin{equation}
n(\vec{p},\vec{Q})=n(\vec{w},-\vec{Q}) \ .  \label{GMD-5-1}
\end{equation}

The quantities $n(\vec{p},\vec{Q})$ and
$\rho_{2h}(\vec{r}_1,\vec{r}_2;\vec{r}_{1'})$ are important
descriptors of nucleon-nucleon correlations in the nuclear medium
representing the next stage of complexity beyond the
single-particle momentum distribution and the one-body density
matrix. Therefore, the last years interest has risen for their
study. One main reason is the need to properly analyze recent and
future experiments of inclusive  character, such as (e,e'), (p,p')
reactions, as well as of exclusive character such as (e,e'N),
(p,2p), ($\gamma$,N) and (e,e'2N), ($\gamma$,2N) and to extract
reliable values for the momentum distribution, the one- and
two-particle spectral functions, the transparency  and other
quantities \cite{Aming-99}-\cite{Jlab}. To achieve these goals,
one must take into account the final-state interactions (FSI) of
the struck nucleons as they propagate through the nuclear medium.
The half-diagonal 2DM appears in almost all quantitative
microscopic post-mean field treatments of the FSI (see for
instance, in the case of inclusive (e,e') scattering
\cite{Petraki-03} and references therein). The GMD, as shown
before, (Eq.~(\ref{GMD-2})), is directly involved in FSI
mechanisms.
Finally, in analogy with other quantum many-body systems
\cite{Magganti-97,Chiafalo-96,Stringari-92}, the functions
$\rho_{2h}(\vec{r}_1,\vec{r}_2;\vec{r}_{1'})$ and
$n(\vec{p},\vec{Q})$ are expected to enter in fundamental sum
rules that furnish insight into the nature of elementary
excitations of the nuclear system.

Initially, both the half-diagonal 2DM and the GMD were
investigated within the context of final-state effects in
inelastic neutron scattering from quantum fluids (i.e. liquid He).
With regard to nuclear systems, calculations based on the work of
Ristig and Clark \cite{Ristig-90} within the context of
variational theory have been performed for Jastrow-correlated
infinite nuclear matter. This has been done for the GMD using
low-order cluster truncations \cite{Clark-93} and later for the
GMD and the half-diagonal 2DM using Fermi-hypernetted-chain
procedures \cite{Mavro-95,Petraki-01}. These results clearly show
the effect of short-range correlations (SRC) in the nuclear medium
and could be used by means of a suitable local density
approximation for the approximate evaluation of the GMD of finite
nuclei. At the same time, the development of rather simple
expressions for the latter which could be easy to use, led to the
study of GMD of closed-shell nuclei within the
independent-particle model and to the extraction of closed
analytical expressions using a harmonic oscillator (HO) basis
\cite{Papa-00}. Only the proton contribution to GMD was considered
and the approach applied was an extension of the one applied in
Refs. \cite{Kosmas-90,Kosmas-92,Papa-98} for the calculation of
the charge form factor, the nuclear charge, matter and momentum
distribution and the one-body density matrix. Recently, the same
approach has been used for the calculation of the two-body
momentum distribution \cite{Papa-03}. The above results for GMD of
nuclei exhibit interesting features rising from the finite size
and the Fermi statistics and are expected to be valid in certain
regions of momenta $p$ and $Q$ where dynamical correlations do not
play a significant role.

In this paper, we extend the above study of GMD in $Z=N$,
$\ell$-closed shell nuclei by considering the proton and neutron
contribution and including Jastrow-type correlations via the
lowest term of a cluster expansion. This so-called low-order
approximation (LOA) \cite{Gaudin-71} has been exploited for the
1DM and 2DM by Bohigas and Stringari \cite{Bohigas-80} and Dal Ri,
Stringari and Bohigas \cite{DalRi-82} and has been used to study
single-particle nuclear properties
\cite{Bohigas-80}-\cite{Moustakidis-00} as well as two-body
nuclear properties \cite{Papa-03,Dimitrova-00,Orlandini-95}.
Recently it has been extended for the case of realistic
interactions and corresponding correlations and applied in the
calculations of the ground-state energies, densities and momentum
distributions of $^{16}$O and $^{40}$Ca
\cite{Arias-97,Alvioli-05}. In our application of LOA to the
evaluation of GMD we have used central single-Gaussian correlation
functions, $f(r)=1-c\erm^{-r^2/\beta^2}$, and we have performed
explicitly the calculation of GMD for $^4$He. Due to its high
central density (almost 8 times nuclear matter density) the
nucleus $^4$He is a particular appropriate system to search for
SRC. We find significant deviations from the independent-particle
model picture due to SRC for rather large values of momenta $p$
and (or) $Q$. We also examine the effect of Center of Mass
corrections and we find that it is quite significant and should be
taken into account. The calculation for heavier $\ell$-closed
nuclei is more complicated. We have developed an approximate
scheme which makes use of the above calculated GMD of $^4$He and
which is likely to be valid for not very large mass number $A$.
Results are extracted for $^{16}$O. It should be emphasized that
the approach presented for the calculation of the GMD using
Jastrow-type correlations is simply an exploratory one, aimed at
guiding realistic calculations which include a full correlation
operator. A similar procedure has been followed in the study of
other quantities including the momentum distribution.

In section 2 a brief outline of the calculation of the GMD for
$Z=N$ $\ell$-closed shell nuclei in the independent-particle shell
model with harmonic oscillator wave functions is given. In section
3 a short presentation is made of our estimation of CM corrections
in the evaluation of GMD. In section 4 the effect of SRC on the
GMD is explored by including Jastrow-type correlations via LOA and
results for the case of $^4$He are extracted. Subsequently, the
approximation for the evaluation of GMD of heavier nuclei that
uses the GMD of $^4$He is presented and it is used for the
evaluation of GMD of $^{16}$O. Finally, in section 5 a summary,
conclusions and hints for further development are given.

\section{GMD of $\ell$-closed nuclei in the harmonic oscillator model}

We consider finite nuclei in their ground state within the
independent-particle model. Using the relation of GMD to the 2DM
in the momentum space
$n_2(\vec{p}_1,\vec{p}_2;\vec{p}_{1'},\vec{p}_{2'})$ (Eq.
(\ref{GMD-3})) and the corresponding expression of the latter for a
system of noninteracting fermions we derive the following
expression
\begin{equation}
n(\vec{p},\vec{Q})=AF(Q^2)n_1(\vec{p},\vec{p}-\vec{Q})+n^{\mathrm{st}}(\vec{p},\vec{Q})\
, \label{GMD-6}
\end{equation}
where $F(Q^2)$ is the elastic form factor, $n_1(\vec{p},\vec{p_{\
}}')$ is the 1DM in momentum  space  and $n^{\rm st}(\vec{p},\vec{Q})$
is the exchange term, arising from the statistical correlations
among the noninteracting fermions generated by the Pauli exclusion
principle. In the case of spin- and isospin-independent
Hamiltonians $n^{\mathrm{st}}(\vec{p},\vec{Q})$ equals
\begin{equation}
n^{\mathrm{st}}(\vec{p},\vec{Q})=-\frac{1}{\nu} \int
n_1(\vec{p},\vec{k}+\vec{Q}) n_1(\vec{k},\vec{p}-\vec{Q}) d^3k \ ,
\label{nst-1}
\end{equation}
($\nu$ the degeneracy due to spin-isospin). The evaluation of the
proton contribution to GMD by means of Eqs. (\ref{GMD-6}),
(\ref{nst-1}) using harmonic-oscillator single-particle states -
but ignoring the Coulomb potential among the protons, the effects
of the center of mass motion and the finite size of the nucleons -
has been presented in detail in Refs. \cite{Papa-00,Papa-98} in
the case of $j$-closed shell nuclei. Here, we present the
expressions in the case of $\ell$-closed shell nuclei, since we
will refer explicitly to the nuclei $^4$He and $^{16}$O. Numerical
results will be presented for $\vec{Q}$ parallel to $\vec{p}$
($\vec{Q}=Q_p\hat{p}$,
$\vec{w}=\vec{p}-\vec{Q}=(p-Q_p)\hat{p}=w_p\hat{p}$) and for
$\vec{Q}$ perpendicular to $\vec{p}$ .
We have for the elastic form factor $F(Q^2)$ \cite{Kosmas-92}
\begin{equation}
F(Q^2)=\frac{1}{Z} \erm^{-Q^2b^2/4}
\sum_{\lambda=0}^{N_{\mathrm{max}}}\theta_{\lambda}(Qb)^{2
\lambda} \ , \label{FF-1}
\end{equation}
where $b$ is the harmonic oscillator parameter,
$N_{\mathrm{max}}=(2n+\ell)_{\mathrm{max}}$ is the number of
energy quanta of the highest occupied proton level, and the
coefficients $\theta_{\lambda}$ are rational numbers varying with
$Z$. Their expressions as well as their values for the ten lowest
$n\ell$ levels are given in App.A of Ref.~\cite{Papa-00}. The 1DM
in the case of $\vec{p'}$ parallel to $\vec{p}$ ($\vec{p'}=p_p'
\hat{p}$) is given by
\begin{equation}
n_1(p,p_p')=\frac{b^3}{\pi^{3/2}}\erm^{-p^2b^2/2} \erm^{-p_p'^2b^2/2}
\sum_{\mu=0}^{N_{\mathrm{max}}}(pb)^{\mu}
\sum_{\mu'=0}^{N_{\mathrm{max}}} (p_p'b)^{\mu'}K_{\mu \mu'} \ .
\label{n1-1}
\end{equation}
The coefficients $K_{\mu \mu'}$ are discussed in the App. B of
Ref. \cite{Papa-00}. From expression (\ref{n1-1}) the expression
of the spherically symmetric nucleon  momentum distribution $n(p)$
is obtained \cite{Kosmas-90}
\begin{equation}
n(\vec{p})=n(p)=\frac{b^3}{\pi^{3/2}} \erm^{-p^2b^2}
\sum_{\lambda=0}^{N_{\mathrm{max}}} (pb)^{2\lambda} 2f_{\lambda} \ .
\label{Np-1}
\end{equation}
Inserting expression (\ref{n1-1}) in Eq. (\ref{nst-1}) the
corresponding expression for the exchange term $n^{\mathrm{st}}$ is derived
\begin{eqnarray}
n^{\mathrm{st}}(p, Q_p)&=&-\frac{b^3}{\pi^{3/2}} \erm^{-p^2b^2/2}
\erm^{-w_p^2b^2/2} \erm^{-Q_p^2b^2/4}
\sum_{\mu=0}^{N_{\mathrm{max}}}(pb)^{\mu}
\nonumber \\
 &&\sum_{\mu '=0}^{N_{\mathrm{max}}}(w_pb)^{\mu '}
\sum_{\rho=0}^{2N_{\mathrm{max}}} (Q_pb)^{\rho} C_{\mu \mu' \rho}
\ . \label{nst-2}
\end{eqnarray}
The coefficients $C_{\mu \mu' \rho}$ are equal to zero if $\mu+
\mu' +\rho=$odd and are discussed in App.C of Ref. \cite{Papa-00}.
Using the values listed there one can calculate the GMD of
$\ell$-closed shell nuclei up to $Z$ or $N$ equal to 40.
Expressions for $n_1$ and $n^{\mathrm{st}}$ for $\vec{Q}$ not
parallel to $\vec{p}$ can be found in Refs.
\cite{Papa-00,Papa-98}.

The GMD calculated via the above expressions in the case of
$^4$He, $^{16}$O and $^{40}$Ca, as a function of $Q_p$ for given
$p$, exhibits a bump centered at $Q_p=0$ for $p=0$ and shifted to
higher values of $Q_p$ for $p>0$. A negative part in the GMD
appears at positive $Q_p$, for nuclei heavier than $^4$He, arising mainly from the term $n^{\rm
st}(p,Q_p)$.
A comparison with respective results for an ideal Fermi gas and
for nuclear matter shows that the positive bump and negative part
are bulk properties of the GMD due to Fermi statistics
\cite{Papa-00}.

\section{Center of Mass Corrections}

As mentioned in the Introduction, special care should be taken
when dealing with finite self-bound systems like nuclei,
especially small ones like $^4$He. The CM motion in such cases
cannot be ignored. The wave functions which are used in the
independent particle model (but even in theories which take also
dynamical correlations into account e.g. Brueckner-Hartree Fock,
Variational Monte Carlo) satisfy the Pauli principle but not the
translation invariance. As a consequence, they contain spurious
components which result from the motion of the CM in a non free
state. Effects from these (also know as CM correlations) are found
in the calculation of almost every observable and make impossible
to extract information for the intrinsic properties of nuclei
directly from experimental data. We will follow Ref.
\cite{Shebeko} for the evaluation of CM effects on the GMD (a
brief history of the CM problem is found there). In Ref.
\cite{Shebeko} the two-body density matrix and the two-body
momentum distribution have been studied by using the fixed CM
approximation to construct the intrinsic wave functions, the
Jacobi variables to define the corresponding intrinsic operators
and an algebraic technique based upon the Cartesian representation
of the coordinate and momentum operators to evaluate the
expectation values involved. We proceed accordingly for the
evaluation of the intrinsic GMD.

The momenta in Eqs. (\ref{GMD-1})-(\ref{GMD-4}) should be
interpreted as momenta with respect to the center of mass of the
system. The two-body density matrix, such as that used in Eq.
(\ref{GMD-3}), should describe the transition matrix element of
two nucleons from intrinsic momenta
\[ \vec{p}_A - \vec{P}/A = \vec{p}, \quad  \vec{p}_{A-1} - \vec{P}/A = \vec{k}
\]
to respective intrinsic momenta $\vec{p'}$, $\vec{k'}$. With
$\vec{p}_i$ we denote the momentum of the $i-$th nucleon with
respect to the artificial mean-field center and with
$\vec{P}=\sum_{i=1}^{A}\vec{p}_i$ the CM momentum. Following the
notation of Ref. \cite{Shebeko} we write the relevant operator in
the form (hats denote operators)
\begin{eqnarray}
\hat{n}_2(\vec{p}, \vec{k}; \vec{p'}, \vec{k'}) & = & A(A-1)
  | \hat{\vec{p}}_{A} - \hat{\vec{P}}/A = \vec{p} \rangle
     \langle \hat{\vec{p}}_{A} - \hat{\vec{P}}/A = \vec{p'} |
\nonumber \\
  & &   \otimes
  | \hat{\vec{p}}_{A-1} - \hat{\vec{P}}/A = \vec{k} \rangle
     \langle \hat{\vec{p}}_{A-1} - \hat{\vec{P}}/A = \vec{k'} |
. \label{En21}
\end{eqnarray}
Switching to Jacobi momenta, as defined in Ref.~\cite{Shebeko},
we have
\begin{eqnarray}
\hat{n}_2(\vec{p}, \vec{k}; \vec{p'}, \vec{k'}) & = & A(A-1)
  | \hat{\vec{\eta}}_{A-1} = \vec{p} \rangle
     \langle \hat{\vec{\eta}}_{A-1} = \vec{p'} |
\nonumber \\
 & &   \otimes
  | \hat{\vec{\eta}}_{A-2} = \vec{k} + \frac{1}{A-1}\vec{p} \rangle
     \langle \hat{\vec{\eta}}_{A-2} = \vec{k'} + \frac{1}{A-1}\vec{p'} |
. \label{En21j}
\end{eqnarray}
In Ref. \cite{Shebeko} a different intrinsic TBMD operator
$\hat{n}^{[2]}$ was defined formally in terms of Jacobi momenta,
with a somewhat different physical interpretation. The two
operators, $\hat{n}_2$ and $n^{[2]}$, are connected via the
relation
\begin{equation}
\hat{n}_2(\vec{p}, \vec{k}; \vec{p'}, \vec{k'}) =
\hat{n}^{[2]}(\vec{p}, \vec{k} + \frac{1}{A-1}\vec{p}; \vec{p'},
\vec{k'} + \frac{1}{A-1}\vec{p'}) \label{En212} .
\end{equation}
Notice that both operators depend only on intrinsic (Jacobi)
variables. Since in the simple HO model the CM wave function
factorizes into an intrinsic and a CM part, there is no need to
project out the CM component of the wave function (by means of the
fixed-CM or other method).

The matrix elements of the operator $\hat{n}_2$ can be calculated
using the formalism of Ref. \cite{Shebeko}. For $^4$He, i.e., in
the state $|0s^4\rangle$, its expectation value is given by
\begin{equation}
{n}_2(\vec{p}, \vec{k}; \vec{p'}, \vec{k'})
 = \frac{12b^6}{\pi^3} \left( \frac{A}{A-2} \right)^{3/2}
   \mathrm{e}^{-\frac{A-1}{A-2}\frac{p^2 + {p'}^2 + k^2 + {k'}^2}{2} b^2 }
   \mathrm{e}^{-\frac{1}{A-2} (\vec{p}\cdot\vec{k} + \vec{p'}\cdot\vec{k'}) b^2}
 \label{En21he}
.
\end{equation}

From $n_2(\vec{p}, \vec{k}; \vec{p'}, \vec{k'}) $ we can evaluate
the generalized momentum distribution $n(\vec{p},\vec{Q})$ by
setting $\vec{p'}=\vec{p}-\vec{Q}$ and $\vec{k'} =
\vec{k}+\vec{Q}$ and integrating over $\vec{k}$, according to
Eq.(3). We find
\begin{equation}
n(\vec{p},\vec{Q}) = \frac{12b^3}{\pi^{3/2}} \left( \frac{A}{A-1}
\right)^{3/2} \mathrm{e}^{-\frac{A}{A-1}(p^2 -
\vec{p}\cdot\vec{Q}) b^2} \mathrm{e}^{-\frac{1}{4}\frac{3A-2}{A-1}
Q^2 b^2} \label{GMD-CMC-1} ,
\end{equation}
which can be written also in terms of $\vec{w}=\vec{p}-\vec{Q}$ as
\begin{equation}
n(\vec{p},\vec{Q}) = \frac{12b^3}{\pi^{3/2}} \left( \frac{A}{A-1}
\right)^{3/2} \mathrm{e}^{-\frac{1}{2}\frac{A}{A-1}(p^2 + w^2 )
b^2} \mathrm{e}^{-\frac{1}{4}\frac{A-2}{A-1} Q^2 b^2}
\label{GMD-CMC-2} .
\end{equation}

The corresponding expression for the intrinsic momentum
distribution $n(\vec{p})$ is \cite{Shebeko}
\begin{equation}
n(\vec{p})=\left( \frac{A}{A-1}\right)^{3/2}\frac{4b^3}{\pi^{3/2}}
e^{-\frac{A}{A-1}p^2} \label{mom-d-1}
\end{equation}
The generalized momentum distribution and the momentum
distribution of $^4$He in the HO model without CM corrections
denoted by $n^0(\vec{p},\vec{Q};b)$ and $n^0(\vec{p};b)$
respectively are given by (see Eqs.(12), (11))
\begin{equation}
n^0(\vec{p},\vec{Q};b)=\frac{12b^3}{\pi^{3/2}}e^{-p^2b^2/2}e^{-w^2b^2/2}
e^{-Q^2b^2/4} \label{MD-HO-HE}
\end{equation}
\begin{equation}
n^0(\vec{p};b)=\frac{4b^3}{\pi^{3/2}}e^{-p^2b^2}
.\label{GMD-MD-HO}
\end{equation}
We have written in Eqs. (\ref{En21he})-(\ref{mom-d-1}) the
different coefficients in terms of $A$ and not 4 to point out a
trend in A-dependence of the effect of CM motion.

The variables $p^2$ and $w^2$ appear to scale due to the CM
correlations by the factor $(1-\frac{1}{A})^{-1}$,
 i.e. the inverse Tassie-Barker factor (TBF),
just like the variable $p^2$ in the momentum distribution
\cite{Shebeko,Shebeko-07}. The variable $Q^2$ scales by the factor
$(1-\frac{1}{A-1})$.
These observations can provide an approximate scheme for
introducing CM corrections to the GMD of $^4$He also when SRC
corrections are considered, as we discuss in the next section.

In heavier nuclei, CM corrections should be less important. To
illustrate this, we note that the GMD in the simple
harmonic-oscillator model for any nucleus is given by the same
exponential as above, times a polynomial -- see Eq. (\ref{nst-2})
and Ref. \cite{Papa-00}. After including CM corrections, the
exponentials are expected --  from analytic arguments -- to be
modified by the same $A-$dependent factors as those of $^4$He.
Already for $^{16}$O these factors do not deviate much from unity,
namely $(1-\frac{1}{A})^{-1}=\frac{16}{15}$,
$(1-\frac{1}{A-1})=\frac{14}{15}$.

\section{The Effect of Short-Range Correlations on GMD}

As it has been mentioned above, the evaluation of GMD within the
independent-particle model is expected to be valid in certain
regions of momenta $p$ and $Q$ where dynamical correlations do not
play a significant role. The next step is to consider dynamical,
short-range correlations on the GMD. In this section we use
state-independent central (Jastrow) correlation functions to
introduce the SRC.

\subsection{Low-order approximation for GMD-Application to $^4$He}

\subsubsection{Low-order approximation for GMD}

Our approach is based on the Jastrow formalism and employs the
low-order approximation  (LOA) of Refs.
\cite{Gaudin-71,Bohigas-80,DalRi-82} for the 2DM. Performing the
spin-isospin summation ($\nu=4$) in Eq.(14) of Ref.
\cite{DalRi-82}, we obtain for the correlated 2DM
$\rho_{2}(\vec{r}_1,\vec{r}_2;\vec{r}_{1'},\vec{r}_{2'})$ the
following expression
\begin{eqnarray}
\rho_{2}^{\mathrm{LOA}}(\vec{r}_1,\vec{r}_2;\vec{r}_{1'},\vec{r}_{2'})
&=&[1+g(r_{12},r_{1'2'})]\rho_2^{0}(\vec{r}_1,\vec{r}_2;\vec{r}_{1'},\vec{r}_{2'})\nonumber\\
&+&
\int[g(r_{13},r_{1'3})+g(r_{23},r_{2'3})][\rho_1^0(\vec{r_1},\vec{r}_{1'})
\rho_{2h}^{0}(\vec{r}_2,\vec{r}_3;\vec{r}_{2'}) \nonumber\\
&-& \nu^{-1} \rho_1^0(\vec{r_1},\vec{r}_{2'})
\rho_{2h}^{0}(\vec{r}_2,\vec{r}_3;\vec{r_{1'}}) \nonumber \\
&-& \nu^{-1} \rho_1^0(\vec{r_1},\vec{r_3})
\rho_2^{0}(\vec{r}_2,\vec{r}_3;\vec{r}_{2'},\vec{r}_{1'})] \drm^3r_3
\nonumber \\
&-&\nu^{-1} \int \int g(r_{34},r_{34})
\{\rho_{2h}^{0}(\vec{r}_2,\vec{r}_4;\vec{r}_{3})
\rho_2^{0}(\vec{r}_1,\vec{r}_3;\vec{r}_{1'},\vec{r}_{2'})  \nonumber\\
&+& \rho_1^0(\vec{r_1},\vec{r_3})[\rho_1^0(\vec{r_2},\vec{r}_{2'})
\rho_{2h}^{0}(\vec{r}_3,\vec{r}_4;\vec{r}_{1'}) \nonumber \\
&-&\nu^{-1} \rho_1^0(\vec{r_2},\vec{r}_{1'})
\rho_{2h}^{0}(\vec{r}_3,\vec{r}_4;\vec{r}_{2'}) \nonumber \\
&-&  \nu^{-1} \rho_1^0(\vec{r_2},\vec{r_4})
\rho_2^{0}(\vec{r}_3,\vec{r}_4;\vec{r}_{1'},\vec{r}_{2'})]\} \drm^3
r_3 \drm^3r_4 \ , \label{LOA-1}
\end{eqnarray}
where $r_{ij}=|\vec{r_i}-\vec{r_j}|$, $g(r,r')=f(r)f(r')-1$ and
$f(r)$ is the Jastrow correlation function, which has to obey the
conditions $f(0)<1$ and $f(r)\rightarrow 1$ for $r \rightarrow
\infty$. In Eq.~(\ref{LOA-1}) the uncorrelated 1DM
$\rho_1^0(\vec{r},\vec{r_{\ }}')$, the 2DM
$\rho_2^{0}(\vec{r_1},\vec{r_2};\vec{r}_{1'},\vec{r}_{2'})$ and
the half-diagonal 2DM
$\rho_{2h}^{0}(\vec{r_1},\vec{r}_{2};\vec{r}_{1'})$ are calculated
in the harmonic-oscillator model. The correlated GMD
$n^{\mathrm{LOA}}(\vec{p},\vec{Q})$, is then calculated by Fourier
transforming according to Eq. (\ref{GMD-4}). The correlated
momentum distribution $n^{\mathrm{LOA}}(\vec{p})$ is calculated
likewise in LOA by performing the spin-isospin summation in
expression (13) of Ref. \cite{DalRi-82} for the one-body density
matrix and by Fourier transforming with respect to
$\vec{r_1}-\vec{r}_{1'}$. Since the LOA preserves the
normalization of the density matrices, the GMD calculated in the
LOA obeys the sequential relation (\ref{GMD-5}) if also
$n(\vec{p})$ is calculated within LOA.

\subsubsection{Application to $^4$He}

The calculation has been carried out for the GMD in the nucleus
$^4$He. The momentum distribution $n^0(\vec{p})$ and the GMD
$n^0(\vec{p},\vec{Q};b)$ in the harmonic-oscillator model are
given by Eqs. (\ref{GMD-MD-HO}) and (\ref{MD-HO-HE}) respectively.
The HO parameter $b=1.382$ fm reproduces the experimental value of
the charge root mean square radius (rms) of $^4$He, $\langle
r_{\mathrm{ch},\mathrm{exp}}^2 \rangle^{1/2}=1.67$ fm
\cite{Unes-87} in this model if corrections due to the
center-of-mass motion and finite nucleon size are taken into
account. The evaluation of the momentum distribution and of GMD in
LOA has been carried out using a single-Gaussian correlation
function, $f(r)=1-c\erm^{-r^2/\beta^2}$. The expression of the GMD
so obtained is
\begin{equation}
n^{\rm LOA}(\vec{p},\vec{Q})=n^0(\vec{p},\vec{Q};b)+\frac{12
b^3}{\pi^{3/2}}\left[n^{(1)}(\vec{p},\vec{Q};b,c,y)+
n^{(2)}(\vec{p},\vec{Q};b,c,y)\right] \ , \label{GMD-Cor-1}
\end{equation}
where $y\equiv b^2/\beta^2$ and
\begin{eqnarray}
n^{(1)}(\vec{p},\vec{Q};b,c,y)&=&\frac{c^2}{[(1+2y)(1+4y)]^{3/2}}
\erm^{-\frac{1}{2}\frac{p^2b^2+w^2b^2}{1+2y}}
\erm^{-\frac{1}{4}\frac{Q^2b^2}{(1+2y)(1+4y)}}\nonumber  \\
&-&\frac{c}{(1+3y)^{3/2}}
\erm^{-\frac{1}{2}\frac{1+2y}{1+3y}(p^2b^2+w^2b^2)}
\erm^{-\frac{1}{4}\frac{1+4y}{1+3y}Q^2b^2}
\erm^{+\frac{2y}{1+3y}\vec{p} \cdot \vec{Q}b^2}\nonumber \\
&-&\frac{c}{(1+3y)^{3/2}}
e^{-\frac{1}{2}\frac{1+2y}{1+3y}(p^2b^2+w^2b^2)}
\erm^{-\frac{1}{4}\frac{1+4y}{1+3y}Q^2b^2}
\erm^{-\frac{2y}{1+3y}\vec{w} \cdot \vec{Q}b^2}
\nonumber \\
%
n^{(2)}(\vec{p},\vec{Q};b,c,y)&=&\left[\frac{10c}{(1+2y)^{3/2}}-
\frac{5c^2}{(1+4y)^{3/2}}\right]
\erm^{-\frac{1}{2}(p^2b^2+w^2b^2)} \erm^{-\frac{Q^2b^2}{4}}\nonumber \\
&+&
\frac{2c^2}{[(1+2y)(1+4y)]^{3/2}}\erm^{-\frac{1}{2}\frac{p^2b^2+w^2b^2}{1+2y}}
\erm^{-\frac{1+6y+12y^2}{(1+2y)(1+4y)}\frac{Q^2b^2}{4}} \nonumber \\
&-& \frac{2c}{(1+3y)^{3/2}}
\erm^{-\frac{1}{2}\frac{1+2y}{1+3y}(p^2b^2+w^2b^2)}
\erm^{-\frac{1}{4}\frac{1+y}{1+3y}Q^2b^2}
\erm^{-\frac{y}{1+3y}\vec{p} \cdot \vec{Q} b^2} \nonumber \\
&-&\frac{2c}{(1+3y)^{3/2}}
\erm^{-\frac{1}{2}\frac{1+2y}{1+3y}(p^2b^2+w^2b^2)}
\erm^{-\frac{1}{4}\frac{1+y}{1+3y}Q^2b^2}
\erm^{\frac{y}{1+3y}\vec{w} \cdot \vec{Q} b^2 } \nonumber \\
&+&\frac{2c^2}{(1+4y)^{3/2}} \erm^{-\frac{1}{2}(p^2b^2+w^2b^2)}
\erm^{-\frac{1}{4}\frac{1+2y}{1+4y}Q^2b^2} \nonumber\\
&-&\frac{4c}{(1+2y)^{3/2}} \erm^{-\frac{1}{2}(p^2b^2+w^2b^2)}
\erm^{-\frac{1}{4}\frac{1+y}{1+2y}Q^2b^2} . \label{Enpq2term}
\end{eqnarray}
We realize that the $n(\vec{p},\vec{Q})$ evaluated in LOA fulfils
the properties (\ref{GMD-5}), (\ref{GMD-5-1}). The corresponding
expression for the momentum distribution $n^{\rm LOA}(\vec{p})$ of
$^4$He \cite{Shebeko-07,Moustakidis-00} is
\begin{eqnarray}
n^{\rm LOA}(\vec{p})&=&n^0(\vec{p};b)+\frac{4
b^3}{\pi^{3/2}}\left[ \frac{c^2
\erm^{-p^2b^2\frac{1}{1+2y}}}{[(1+2y)(1+4y)]^{3/2}}- \frac{2c
\erm^{-p^2b^2\frac{1+2y}{1+3y}}}{(1+3y)^{3/2}}\right] \nonumber \\
&+&\frac{4 b^3}{\pi^{3/2}} \left[\left(\frac{6c}{(1+2y)^{3/2}}-
\frac{3c^2}{(1+4y)^{3/2}}\right)\erm^{-p^2b^2}\right. \nonumber
\\
&+& \left.\frac{2c^2
\erm^{-p^2b^2\frac{1}{1+2y}}}{[(1+2y)(1+4y)]^{3/2}}- \frac{4c
\erm^{-p^2b^2\frac{1+2y}{1+3y}}}{(1+3y)^{3/2}}\right].
\end{eqnarray}
The term $n^{(2)}$, Eqs.(\ref{GMD-Cor-1}),(\ref{Enpq2term}),  does not
contribute if only the first two terms of the LOA approximation
are used. This latter case has been considered in Ref.
\cite{Mavro-01}.

The inclusion of CM corrections to the correlated GMD is a rather
tedious task. Only recently, CM and short-range correlations
(within the LOA) have been considered simultaneously, for one-body
quantities, namely the density, form factor and momentum
distribution of $^4$He \cite{Shebeko-07}. In particular, the
corresponding expression for $n(\vec{p})$ if the correlation
function  $f(r)=1-c\erm^{-\beta^2r^2}$ is used for $c=1$ is
\begin{eqnarray}
n^{LOA+CM}(\vec{p})&=&\frac{4b^3}{\pi^{3/2}}\left[\left(1+\frac{6}{(1+2y)^{3/2}}-\frac{3}{(1+4y)^{3/2}}\right)
(\frac{4}{3})^{3/2}\erm^{-\frac{4}{3}p^2b^2}\right. \\
&+&\left.  \frac{24}{[(1+4y)(3+8y)]^{3/2}}
\erm^{-\frac{4p^2b^2}{(3+8y)}}-
\frac{48}{(3+10y)^{3/2}}\erm^{-\frac{4(1+2y)}{(3+10y)}p^2b^2}\right]
\nonumber. \label{Sheb-1}
\end{eqnarray}

For the purposes of the present work we will consider an
approximate scheme to estimate the CM effects on the correlated
GMD of $^4$He, based on the scaling that the CM corrections seem
to introduce to the momentum variables, see Sec. 3. We scale the
variables $\vec{p}$ and $\vec{w}$ by $(\frac{A}{A-1})^{1/2}=
(4/3)^{1/2}$ and the variable $\vec{Q}$ by
$(\frac{A-2}{A-1})^{1/2}=(2/3)^{1/2}$, so that the new GMD obeys
the symmetry property of Eq.~(\ref{GMD-5-1}). In addition we
multiply the resulting expression by $\left( \frac{A}{A-1}
\right)^{3/2}=(4/3)^{3/2}$ to ensure the relation between the GMD
for $\vec{Q}=0$ and the properly normalized momentum distribution
Eq. (\ref{GMD-5}). The latter is obtained from the correlated
momentum distribution, as calculated within the LOA, by applying
the same scaling to the variable $p$ and multiplying the resulting
expression by $\left( \frac{A}{A-1} \right)^{3/2}$. (We will call
the above approximate scheme 'LOA+CM').

Numerical results for GMD have been obtained with two
parametrizations of $f(r)$. First, the Gaussian correlation
function (1G) which has been used in evaluating several quantities
of $^4$He
\cite{Bohigas-80,DalRi-82,Stoitsov-93,Massen-99,Dimitrova-00,Orlandini-95,Papa-03}
has been considered. The values of its parameters $c$ and $\beta$
are  $0.76$ and $0.83$~fm  respectively and the wound parameter
$\kappa$ ( $\kappa =\int(1-f(r_{12}))^2 \rho_2^0(r_{12})
d^3r_{12}$, $\rho_2^0(r_{12})$ is the relative pair density
distribution function calculated in the HO model and normalized to
unity) equals $0.018$. Second, the Gaussian correlation function
($g_2$) which has been used for the calculation of GMD of infinite
nuclear matter with the FHNC/0 approach \cite{Mavro-95},
\cite{Flyn-84} has been used aiming on one hand to compare the LOA
results in $^4$He  with the FHNC/0 results in infinite nuclear
matter obtained with the same correlation function, and on the
other to test the sensitivity of the results on the strength of
the correlation function used. In addition, $g_2$ has been used to
estimate the CM corrections on $n(\vec{p})$, as discussed later.
The involved parameters $c$ and $\beta$ equal to $1$ and $0.6765$
fm respectively. The wound parameter for $g_2$ equals $0.016$. The
two correlation functions are plotted in Fig. 1.


Some numerical results for $n(\vec{p},\vec{Q})$ of $^4$He are
presented in Figs. 2-6. In Fig. 2, the GMD is plotted for
$\vec{Q}$ parallel to $\vec{p}$, $\vec{Q}=Q_p \hat{p}$ as a
function of $Q_p$ for $p=0,1,2$ and $3$ fm$^{-1}$. The correlated
GMD calculated with the use of LOA and of the correlation
functions 1G and $g_2$, as given by Eq. (\ref{GMD-Cor-1}), is
plotted by continuous and dashed lines respectively. The GMD in
the harmonic- oscillator model, Eq. (\ref{MD-HO-HE}), is plotted
by dotted lines. It seems that SRC within LOA contribute a
negative part to $n(\vec{p},\vec{Q})$ for positive $Q_p$ mainly at
high values of $p$. As expected, deviations from the HO picture
are larger for high values of $p$ and (or) $Q_p$. We should recall
that in infinite nuclear matter it was found that SRC are mainly
significant for $|\vec{p}-\vec{Q}|>k_F$ and/or $p>k_F$
\cite{Mavro-95}. Results obtained with correlation functions 1G
and $g_2$ are similar, but one should bear in mind that they imply
about the same strength of SRC (as judged by the corresponding
values of the wound parameter).

Initially the effect of the CM motion in $^4$He have been
considered on the generalized momentum distribution per pair for
$Q=0$ (which equals the momentum distribution per particle (Eq.
(5)).  In Fig.~3 the momentum distribution $n(p)$ per particle has
been plotted in the single HO model (Eq. (21)) and in the HO
including CM effects (HO+CM, Eq. (19)) as well as in LOA including
SRC via correlation function $g_2$ (LOA, Eq. (25)). CM effects
have been evaluated exactly (LOA+CM, Eq. (26)) and using
approximation 'LOA+CM'. We realize that in $^4$He in the LOA+CM
evaluation of $n(p)$ a shrinking is observed relative to its
values in LOA which affects the low-medium range of $p$, while the
high momentum tail is not affected. A similar shrinking occurs in
the HO evaluation. In addition one notices that our approximation
'LOA+CM' is quite satisfactory for estimating  the CM effects on
$n(p)$ and can be used also in the case of $n(\vec{p},\vec{Q})$.
In Figs~4 and 5 the GMD is plotted for $\vec{Q}$ parallel
$\vec{p}$ and for $\vec{Q}$ perpendicular to $\vec{p}$ for this
case, 'LOA+CM' approximation, as well as for LOA approximation
(LOA, Eq. (23)) with correlation function $g_2$. For comparison,
plots are shown for the simple harmonic oscillator (HO, Eq. (20))
and with CM effects (HO+CM, Eq.(17)). We realize that CM effects
modify the range of values of $n(\vec{p},\vec{Q})$ and produce a
similar change in LOA and HO evaluations.

In Fig. 6, a comparison is made for the GMD per particle for
 $\vec{Q}$ parallel to $\vec{p}$ in the case of $^4$He (continuous
line) and infinite nuclear matter (diamond chain) calculated
within the LOA and FHNC/0 approximation \cite{Mavro-95}
respectively. The Fermi wave number $k_F$ of infinite nuclear
matter has been taken equal to $1.3925$ fm$^{-1}$ (density
$\rho^{\rm NM}=0.182$~fm$^{-3}$) and in both systems the
correlation function $g_2$ has been used. Results are shown for
$p$ equal to $0$ and $2k_F^{\rm NM}$. We observe qualitative and
in some cases even quantitative agreement in the range where SRC
are expected to dominate ($p>k_F$ and/or $|\vec{p}-\vec{Q}|>k_F$).
This implies that the convergence of LOA is satisfactory. The
disappearance of discontinuities in the behavior of GMD of $^4$He
at $Q_p=|p-k_F|$ or $Q_p=|p+k_F|$ characterizes the transition
from infinite  to finite Fermi systems.

Regarding the effect of other than central correlations an
estimate can be drawn for the GMD per pair for $Q=0$ (which equals
$n(p)/A$, Eq. (5)). In Fig.~3 the results of calculations of
$n(p)/A$ with variational Monte Carlo method \cite{Pieper-92}
(dashed line) are plotted along with our results. It seems that
the deviations between our results (in LOA approximation) and
those of ref. \cite{Pieper-92} originate mostly from such
correlations.

\subsection{Approximation for heavier nuclei using LOA for $^4$He.
Application to $^{16}$O}

\subsubsection{Approximation for heavier nuclei }

In principle, one could evaluate the effect of SRC on the GMD of
all $\ell$-closed shell $Z=N$ nuclei within the LOA approximation
for the 2DM  (Eq. (\ref{LOA-1})) using the results for the one and
two-body density matrices in the harmonic oscillator model
\cite{Papa-00} and Fourier transforming according to Eq.
(\ref{GMD-4}). The resulting expressions even in the case of the
nucleus $^{16}$O are rather long and complicated. We propose an
approximation for calculating  the GMD for not too heavy nuclei
which makes use of the GMD of $^4$He calculated in LOA. It has
been derived from a corresponding approximation that seems to be
valid for the momentum distribution $n(\vec{p})$. Microscopic
calculations of $n(\vec{p})$ indicate that for large values of $p$
($p>k_F$) the momentum distribution per nucleon is mainly
dominated by the SRC and is almost independent of $A$, whereas for
small values of $p$ it is satisfactorily described within the
independent particle model \cite{Flyn-84,Arias-96,Antonov-05}.
Recently, it has also been shown experimentally that the momentum
distribution at high momenta has the same shape for all nuclei
differing only by a scale factor \cite{Epiyan-03}. Therefore, the
following approximation for $n(\vec{p})$ seems reasonable for a
nucleus of mass number $A$, if we use the momentum distribution
per particle of the nucleus $^4$He ($A=4$)
\begin{equation}
n^{\rm corr-4}(\vec{p};A)=n^0(\vec{p};A)+\frac{A}{4}\Delta n
(\vec{p};4) \ , \label{ncor-He-1}
\end{equation}
where $n^0(\vec{p};A)$ and $n^{\rm corr-4}(\vec{p};A)$ are the
momentum distribution in the in-dependent-particle model and
including SRC respectively and $\Delta n (\vec{p};4)=n^{\rm
corr}(\vec{p};4)-n^0(\vec{p};4))$. Using the sequential relation,
Eq. (\ref{GMD-5}), and the correlated GMD of the nucleus $^4$He,
we can derive the corresponding approximation for GMD of the
nucleus with mass number $A$
\begin{equation}
n^{\rm
corr-4}(\vec{p},\vec{Q};A)=n^{0}(\vec{p},\vec{Q};A)+\frac{A(A-1)}{12}
\Delta n (\vec{p},\vec{Q};4)\ ,  \label{npq-corr}
\end{equation}
where $n^{0}(\vec{p},\vec{Q};A)$ and $n^{\rm
corr-4}(\vec{p},\vec{Q};A)$ are the GMD in the
independent-particle model and including SRC respectively and
$\Delta n(\vec{p},\vec{Q};4)=n^{\rm
corr}(\vec{p},\vec{Q};4)-n^{0}(\vec{p},\vec{Q};4)$. We will use
LOA to evaluate $n^{\rm corr}(\vec{p},\vec{Q};4)$ (Eq. (28)) and
we will call the corresponding expression of $n^{\rm
corr}(\vec{p},\vec{Q};A)$ as LOA-4. The approximation
(\ref{npq-corr}) is valid for not large values of $A$, as it does
not produce a correct asymptotic behaviour for
$A\rightarrow\infty$.

\subsubsection{Application to $^{16}$O}

In this paper, we have applied approximation (\ref{npq-corr}) to
calculate the effect of SRC on the GMD of the nucleus $^{16}$O. We
have considered the special case that $\vec{p}$ and $\vec{Q}$ are
parallel. Results for the GMD of this nucleus within the
harmonic-oscillator model, $n^{0}(\vec{p},\vec{Q};16)$, have been
presented and discussed in Ref. \cite{Papa-00}. As for $\Delta n
(\vec{p},\vec{Q};4)$ we have used our results for the nucleus
$^4$He within the LOA for the correlation function 1G, presented
in section 4.1.2. The harmonic oscillator parameter $b$ for both
nuclei $^4$He and $^{16}$O has been determined in such a way as to
reproduce the experimental value of the charge r.m.s. radius
($\langle r_{\rm ch,exp}^2\rangle ^{1/2}=1.67$ fm and $2.737$ fm
respectively \cite{Unes-87}). We found the values $1.2195$ fm and
$1.7825$ fm respectively within the harmonic-oscillator model.

Some of our results are plotted in Figs.~7 and 8. Fig.~7
illustrates the variation of $n(p,Q_p)$ as a function of $Q_p$ for
$p=0,1.1,2$ and $3$ fm$^{-1}$. Among the values of $p$ considered,
we have included the value $1.1$ fm$^{-1}$, which corresponds to
the Fermi wave number $k_F$ in $^{16}$O \cite{Papa-00,Pieper-92}.
The results obtained with the approximation (\ref{npq-corr})  and
within the HO model are displayed. We realize, as expected, the
effect of SRC mainly for $p>k_F$. Since for the description of SRC
in the GMD of $^{16}$O use is made of the above evaluation of GMD
of $^4$He, the conclusions drawn for their effect on the behavior
of GMD of $^{16}$O are similar. Fig. 8 provides an estimate of the
quality of the approximation (\ref{npq-corr}), of the magnitude of
the omitted higher-order terms and of the contribution of other
than central correlations. The GMD per pair for $Q=0$ calculated
within (\ref{npq-corr}), as described above (which equals the
momentum distribution per particle $n(p)/A$ according to Eq.
(\ref{GMD-5})) is compared with the $n(p)/A$ calculated within
variational Monte Carlo method \cite{Pieper-92} and the
harmonic-oscillator model.
Judging from the quite good agreement between LOA and FHNC results
for the momentum distribution found in ref. \cite{Arias-97} with
the use of the same correlation function and the results presented
in ref. \cite{Alvioli-05} it seems that the major part of the
deviation between our results and those of ref. \cite{Pieper-92}
stems from other than central correlations. As mentioned in Sec.~3
the CM corrections on the GMD of $^{16}$O estimated by using
suitable scaling for $\vec{p}$, $\vec{w}$ and $\vec{Q}$, or a more exact method, are
expected to be small.

\section{Summary and Conclusions}
In summary, the study of the generalized momentum distribution
$n(\vec{p},\vec{Q})$ of finite, $Z=N$, $\ell$-closed shell nuclei
in their ground state that has been started in Ref. \cite{Papa-00}
within the context of the independent-particle shell model using
harmonic-oscillator wave functions, was continued by including
Jastrow-type correlations for investigating the effect of
short-range correlations, which is expected to be important in
certain regions of momenta $p$ and $Q$. First, the low-order
approximation of Ref. \cite{DalRi-82} has been used and the
$n(\vec{p},\vec{Q})$ of $^4$He has been evaluated using
single-Gaussian correlation functions (parametrizations 1G and
$g_2$). Significant deviations from the independent-particle
picture were found for rather large values of $p$ and (or) $Q$
($p>k_F$, $|\vec{p}-\vec{Q}|>k_F$). The convergence of LOA was
explored by comparing with the results of the FHNC/0 calculation
of $n(\vec{p},\vec{Q})$ of infinite nuclear matter \cite{Mavro-95}
in which the same correlation function $g_2$ was used and it was
found satisfactory in most cases. In addition, the effects of CM
motion has been estimated and found not negligible and the role of
correlations other than central has been brought up.
Next, an approximation scheme for the evaluation of
$n(\vec{p},\vec{Q})$ of heavier nuclei was proposed (Eq.
(\ref{npq-corr})) that includes the effect of short-range
correlations by means of the above evaluated GMD of $^4$He.
Numerical results have been derived for $^{16}$O and the quality
of the approximation was discussed.

Further investigation of  $n(\vec{p},\vec{Q})$ of finite nuclei
should consider the exact evaluation of the corrections due to the
Center-of-Mass motion which are quite significant in the case of
light nuclei   as we have realized in the approximate treatment of
$^4$He. One should start along  the lines of references
\cite{Shebeko,Shebeko-07} for the case of $N=Z$, l-closed nuclei.
Starting from these nuclei, one must also consider to include in
the calculations of $n(\vec{p},\vec{Q})$ state-dependent
correlations. Another interesting direction for future work is the
determination of other Fourier transforms of the two-body density
matrix, for example $n(\vec{p},\vec{k},\vec{Q})$
\cite{Ristig-89,Moustakidis}. The above evaluation of
$n(\vec{p},\vec{Q})$ is a first step towards more realistic
calculations which include also other than Jastrow correlations.
The quantity $n(\vec{p},\vec{Q})$ is mainly useful for the study
of final-state interactions of struck nucleons as they propagate
inside the nuclear medium in various scattering processes and for
the understanding of elementary excitations of nuclei.

\section*{Acknowledgments}
 The work of Ch.C. Moustakidis was supported by the
Pythagoras II Research project of EPEAEK (80861) and the European
Union. Partial financial support from the University of Athens
under Grants 70/4/3309 and from the Deutsche
Forschungsgemeinschaft through contract SFB 634 is acknowledged.

\newpage

\begin{figure}
\centering
\includegraphics[height=8.0cm,width=8.0cm]{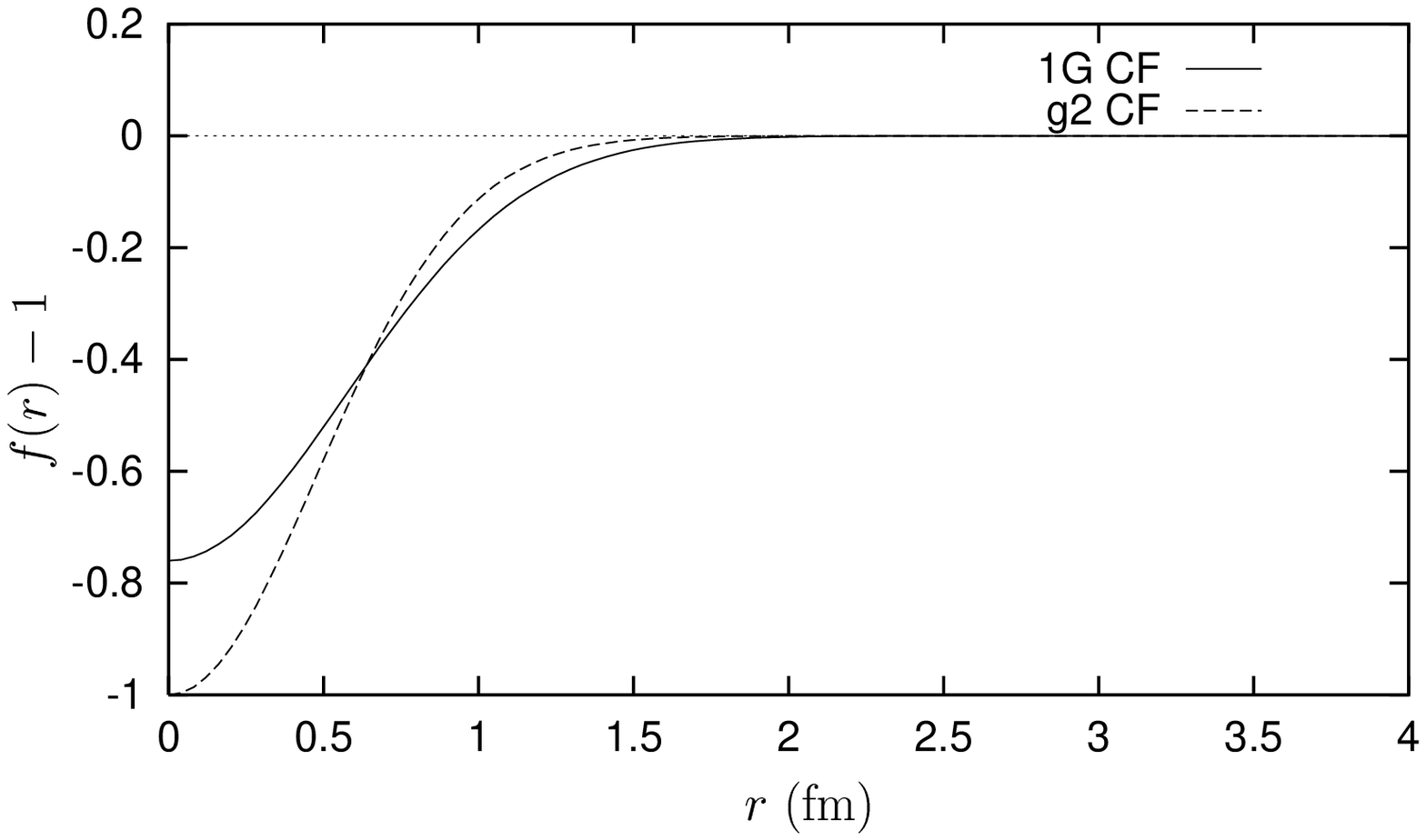}
\caption{The single-Gaussian correlation functions with $c=0.76$,
$\beta=0.83$ fm (1G)(continuous line) and with $c=1$,
$\beta=0.6765$ fm ($g_2$)(dashed line) (see 4.1.2). } \label{}
\end{figure}
\clearpage \clearpage
\begin{figure}
\centering
\includegraphics[height=10.0cm,width=10.0cm]{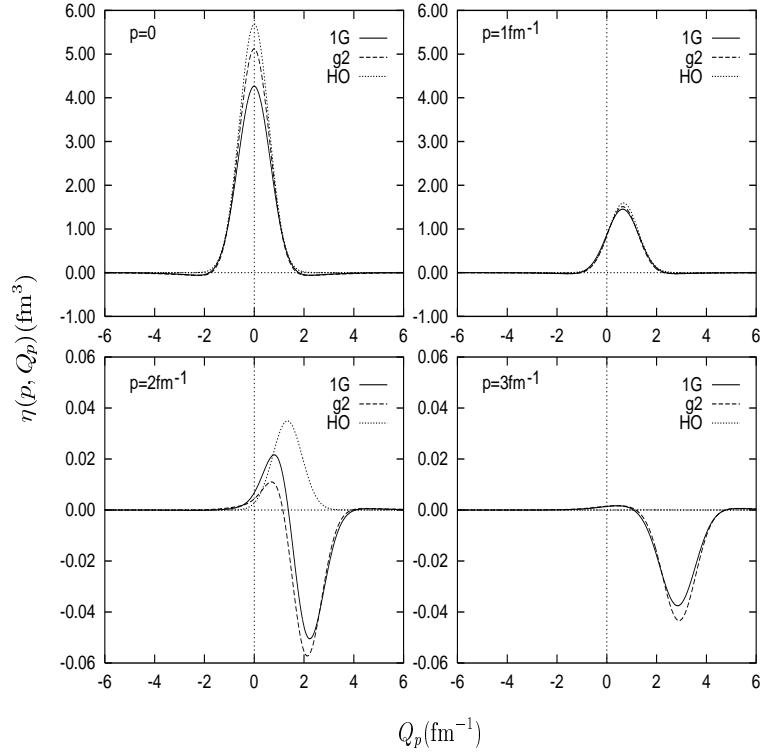}
\caption{Generalized momentum distribution of $^4$He for $\vec{Q}$
parallel to $\vec{p}$, $n(p,Q_p)$ as a function of $Q_p$
 for $p=0,1,2,3$ fm$^{-1}$ including SRC with LOA, Eq.(23) using correlation
 functions 1G (continuous line) and $g_2$ (dashed line) and in the
 harmonic-oscillator model, Eq. (20) (dotted line). (see 4.1.2)} \label{}
\end{figure}
\clearpage \clearpage
\begin{figure}
\centering
\includegraphics[height=10.0cm,width=10.0cm]{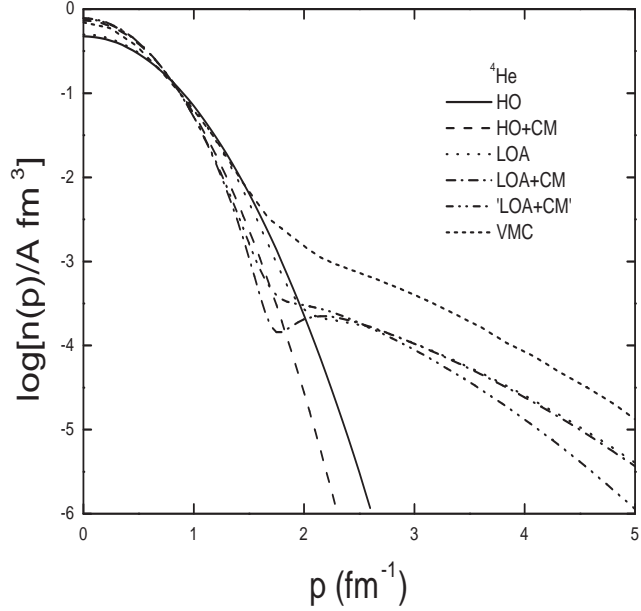}
\caption{The momentum distribution per particle (or the
generalized momentum distribution per pair for $Q=0$ (Eq.(5)) as a
function of $p$ in the case of $^4$He in the simple harmonic
oscillator model (HO, Eq.(21)) and with CM effects (HO+CM, Eq.
(19)) as well including SRC using LOA and correlation function
$g_2$ (LOA, Eq.(25)) and considering CM effects (LOA+CM, Eq.(26))
and approximation 'LOA+CM' (see sec. 4.1.2). In addition, the
results of the variational Monte Carlo calculation VMC [49] are
plotted. A logarithmic scale is used. } \label{}
\end{figure}
\clearpage \clearpage
\begin{figure}
\centering
\includegraphics[height=6.0cm,width=6.0cm]{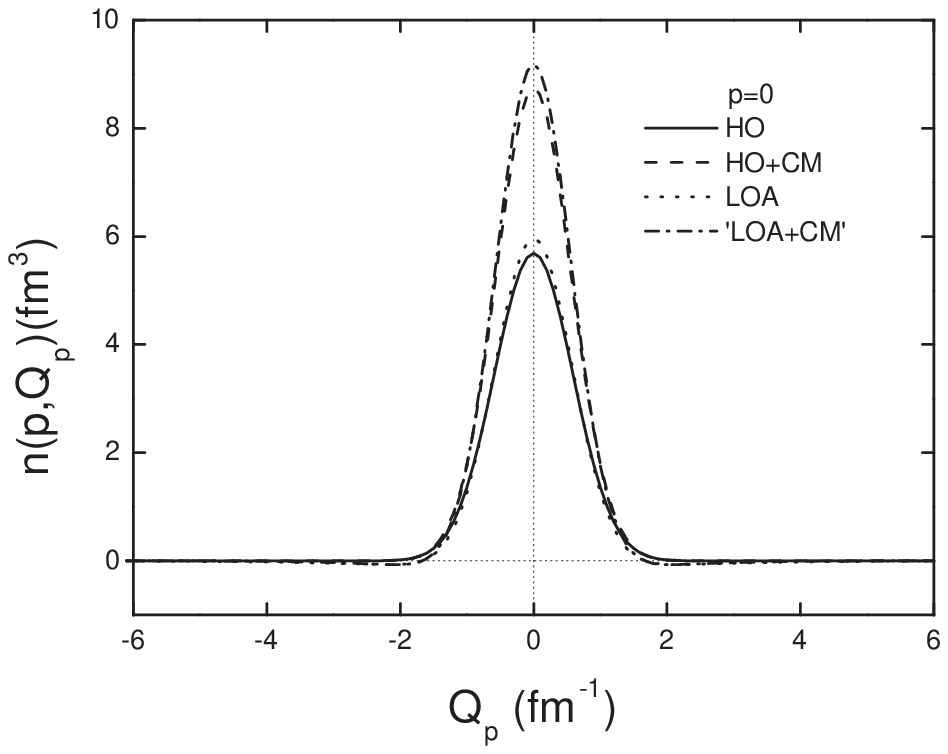}
\includegraphics[height=6.0cm,width=6.0cm]{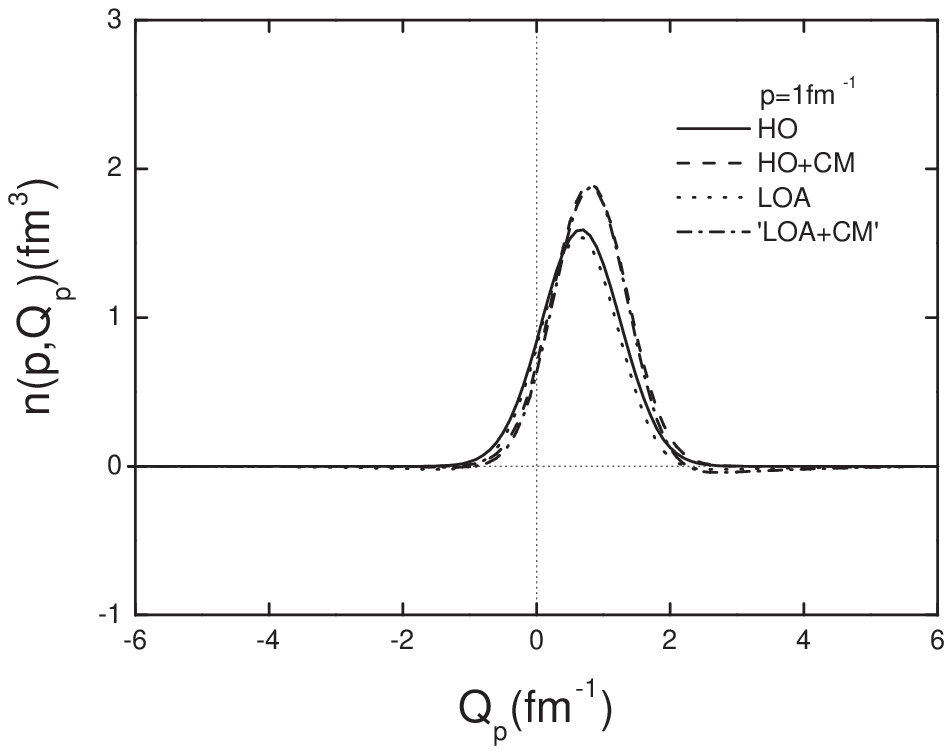}\\
\includegraphics[height=6.0cm,width=6.0cm]{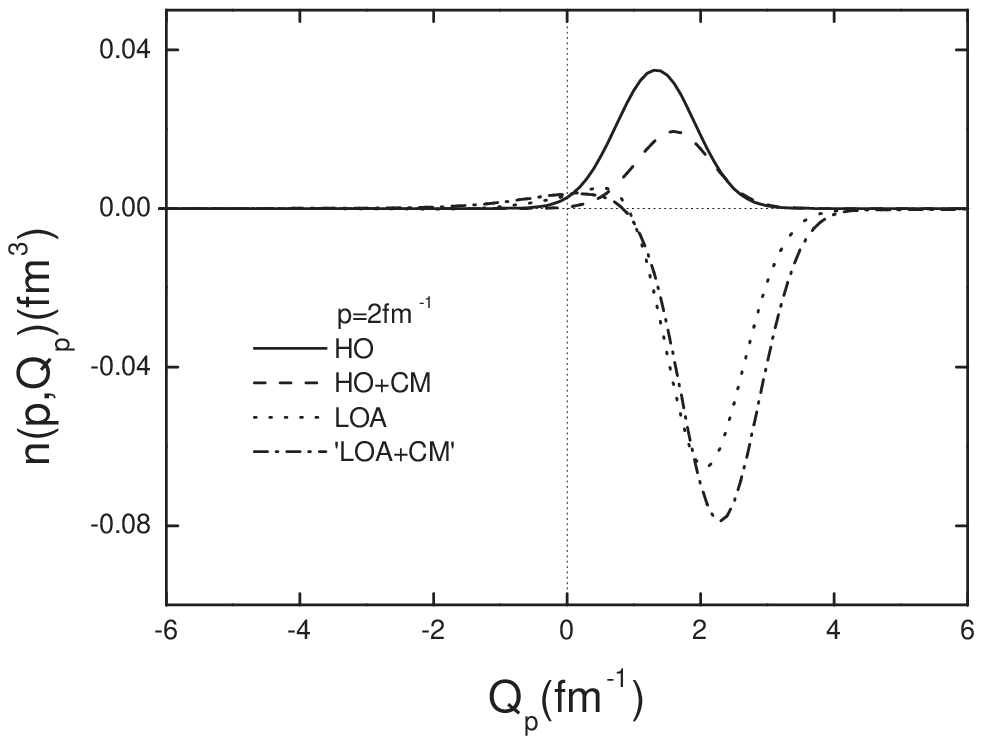}
\includegraphics[height=6.0cm,width=6.0cm]{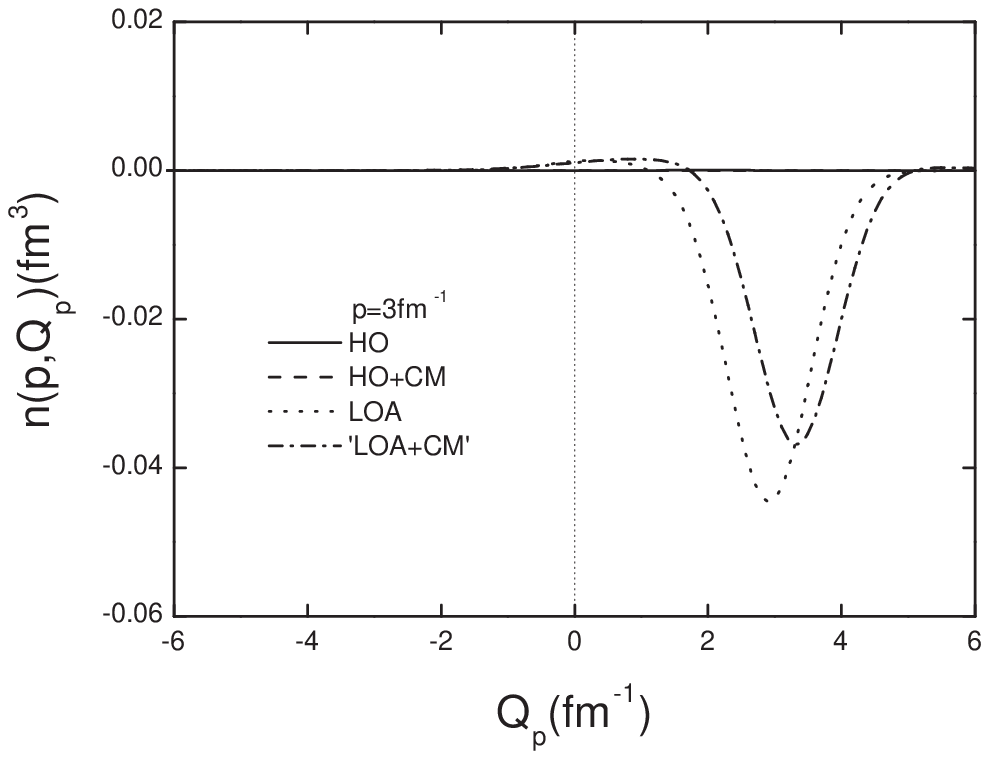}
\caption{Generalized momentum distribution of $^4$He for $\vec{Q}$
parallel to $\vec{p}$, $n(p,Q_p)$ as a function of $Q_p$ for
$p=0,1,2,3$ fm$^{-1}$ in the simple harmonic oscillator (HO,
eq.(20)) and  with CM effects (HO+CM, eq. (17)) as well including
SRC using LOA and correlation function $g_2$ (LOA, eq.(23)) and
considering CM effects using approximation 'LOA+CM' (see 4.1.2).}
\label{}
\end{figure}
\clearpage \clearpage
\begin{figure}
\centering
\includegraphics[height=6.5cm,width=6.5cm]{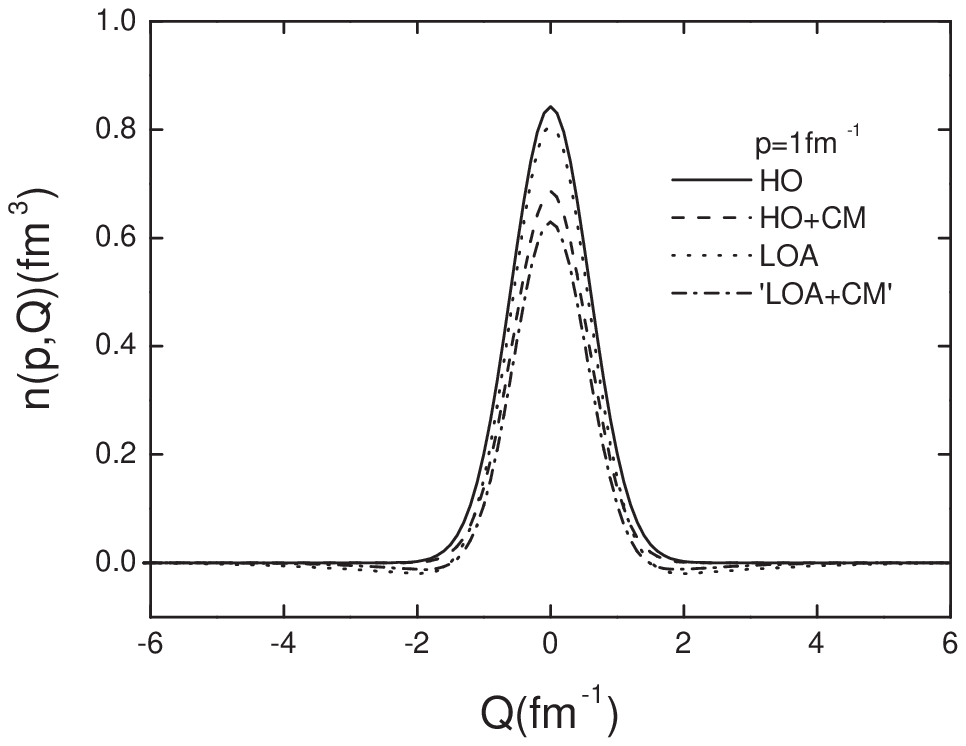}
\includegraphics[height=6.5cm,width=6.5cm]{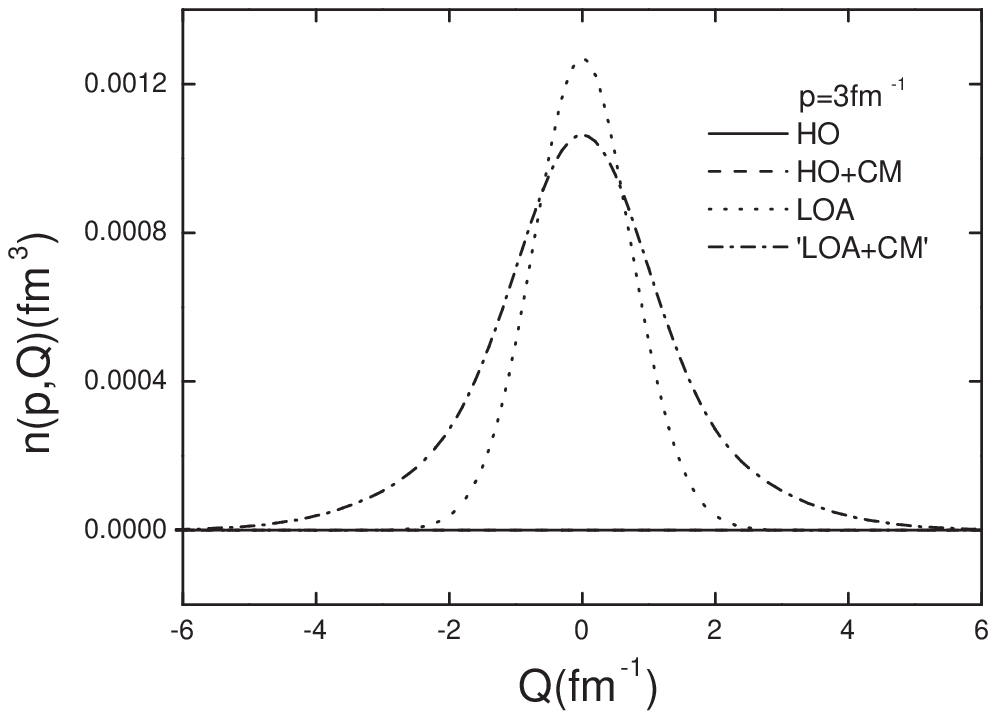}
\caption{As in figure 4 but for $\vec{Q}$ perpendicular to
$\vec{p}$ and for $p=1$ and $p=3$ fm$^{-1}$.} \label{}
\end{figure}
\clearpage \clearpage

\begin{figure}
\centering
\includegraphics[height=10.0cm,width=10.0cm]{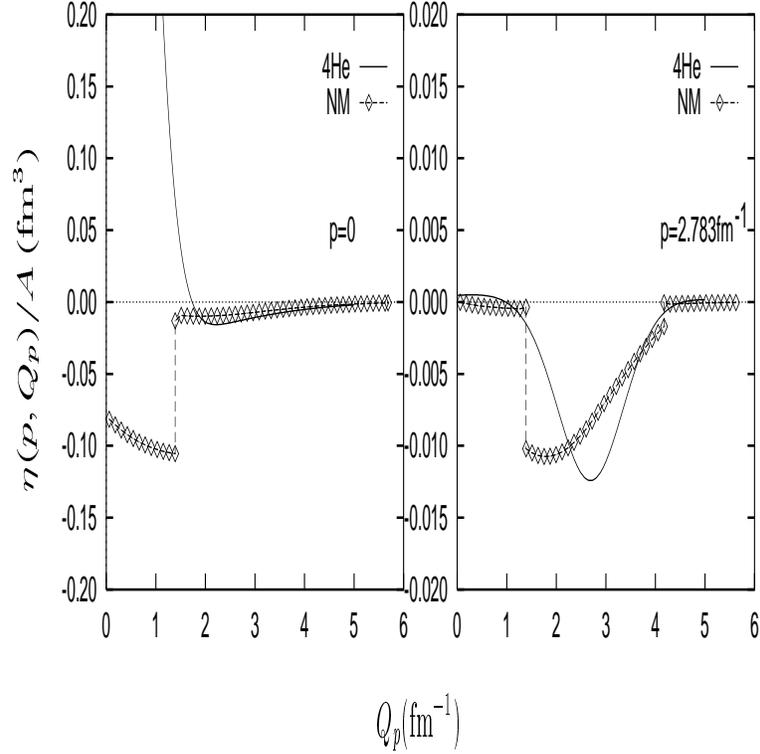}
\caption{Generalized momentum distribution per particle for
$\vec{Q}$ parallel to $\vec{p}$, $n(p,Q_p)/A$ as a function of
$Q_p$ for $^4$He calculated by including SRC using LOA (continuous
line, Eq.(23)) and for infinite nuclear matter with Fermi wave
number $k_F^{NM}=1.3915$ fm$^{-1}$ calculated in Fermi-hypernetted
chain approximation \cite{Mavro-95} (diamond chain). In both
systems the correlation function $g_2$ is used and results are
presented for $p=0$ and $2$ $k_F^{NM}$. } \label{}
\end{figure}
\clearpage \clearpage
\begin{figure}
\centering
\includegraphics[height=10.0cm,width=10.0cm]{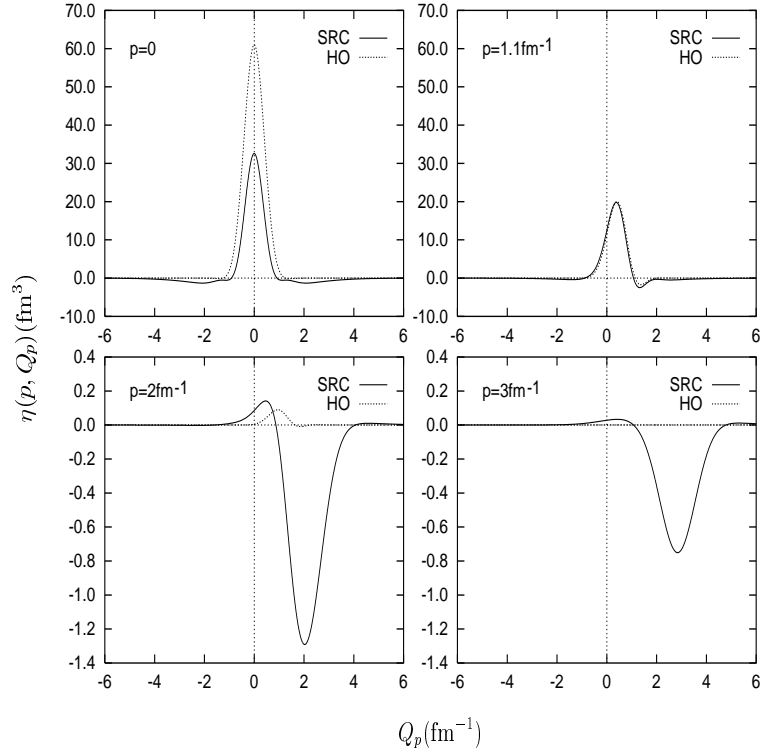}
\caption{Generalized momentum distribution of $^{16}$O for
$\vec{Q}$ parallel to $\vec{p}$, $n(p,Q_p)$ as a function of $Q_p$
for $p=0,1.1,2$ and $3$ fm$^{-1}$ including SRC using the
approximation of Eq.(\ref{npq-corr}) and LOA (LOA-4) and
correlation function 1G (continuous line) and in the
harmonic-oscillator model (dotted line). } \label{}
\end{figure}
\clearpage \clearpage
\begin{figure}
\centering
\includegraphics[height=10.0cm,width=10.0cm]{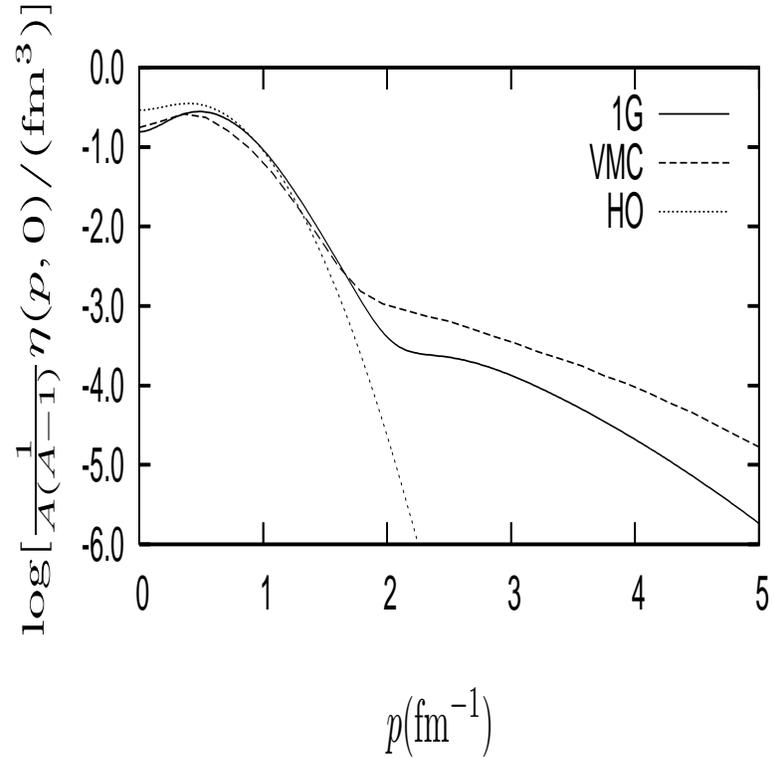}
\caption{Generalized momentum distribution per pair for $Q=0$ (or
momentum distribution per particle (Eq. (5)) as a function of $p$,
in the case of $^{16}$O, including SRC using the approximation of
Eq.(\ref{npq-corr}) and LOA (LOA-4) and correlation function 1G
(continuous line) along with the results for the momentum
distribution per particle of a variational Monte Carlo calculation
\cite{Pieper-92} (dashed line, VMC) and of the harmonic-oscillator
model (dotted line). A logarithmic scale is used. } \label{}
\end{figure}
\clearpage \clearpage

\end{document}